# Coexistence of superconductivity and itinerant ferromagnetism in $Sr_{0.5}Ce_{0.5}FBiS_{2-x}Se_x$ ($x$ = 0.5 and 1.0), the first non-U material with $T_c < T_{FM}$

Gohil S. Thakur[1], G. Fuchs[2], K. Nenkov[2], V. Grinenko[2], Zeba Haque[1], L. C. Gupta[1‡], and A. K. Ganguli[1,3]*

[1]Department of Chemistry, Indian Institute of Technology, New Delhi, India, 110016

[2]Leibniz-Institut für Festkörper- und Werkstoffforschung Dresden, Germany, 01069

[3]Institute of Nano Science and Technology, Mohali, Punjab, India, 160064

## Abstract

We have carried out detailed magnetic and transport studies of the *new* $Sr_{0.5}Ce_{0.5}FBiS_{2-x}Se_x$ ($x$ = 0.5, 1) superconductors derived by doping Se in $Sr_{0.5}Ce_{0.5}FBiS_2$. Se–doping produces several effects: it suppresses semiconducting–like behavior observed in the undoped $Sr_{0.5}Ce_{0.5}FBiS_2$, ferromagnetic ordering temperature, $T_{FM}$, decreases considerably from 7.5 K (in $Sr_{0.5}Ce_{0.5}FBiS_2$) to 3.5 K and superconducting transition temperature, $T_c$, gets enhanced slightly to 2.9 – 3.3 K. Thus in these Se–doped materials, $T_{FM}$ is just marginally *higher* than $T_c$. Magnetization studies provide an evidence of bulk superconductivity in $Sr_{0.5}Ce_{0.5}FBiS_{2-x}Se_x$. Quite remarkably, as compared with the effective paramagnetic Ce–moment (~ 2.2 $\mu_B$), the ferromagnetically ordered Ce–moment in the superconducting state is *rather* small (~ 0.1 $\mu_B$). To the best of our knowledge, the title compounds are the *first Ce–based superconducting itinerant ferromagnetic materials* ($\boldsymbol{T_c < T_{FM}}$). We stress that Ce–4*f* electrons are responsible for **both** superconductivity and ferromagnetism just as *U–5f* electrons are in UCoGe. Furthermore, a novel feature of these materials *is a* ***dual*** *hysteresis loop corresponding to* ***both*** *the ferromagnetism and the coexisting superconductivity. Such features of* $Sr_{0.5}Ce_{0.5}FBiS_{2-x}Se_x$ put these materials *apart from the well known U–containing superconducting ferromagnets reported so far.*

## Introduction

Traditionally, superconductivity and long range ferromagnetism had been considered mutually exclusive (BCS pair–breaking and Meissner effect). Over the last several years a number of materials that exhibit coexistence of superconductivity and long range ferromagnetism have been discovered. In materials such as $ErNi_2B_2C$ [1,2] and $RuSr_2GdCu_2O_8$ [3], localized 4*f*–moments (Er, Gd) are responsible for long range ferromagnetism whereas 3*d*–conduction electrons carry superconductivity. On the other hand, in the $UGe_2$,[4] URhGe [5], UIr [6], UCoGe [7], U–5*f* itinerant electrons are responsible both for superconductivity and magnetism. The material UCoGe is of particular interest from the view point of the present work. In this material the paramagnetic effective moment of U is ~ 1.7 $\mu_B$ whereas the ferromagnetic ordered moment of U is drastically reduced, 0.03 $\mu_B$[7]. The materials under investigation (*the title compounds*) in this work, exhibit coexisting superconductivity and itinerant ferromagnetic properties, as we shall see below, similar to those of UCoGe. To the best of our knowledge, $Sr_{0.5}Ce_{0.5}FBiS_{2-x}Se_x$ is the *first Ce-containing materials exhibiting coexisting superconductivity and itinerant ferromagnetism.*

Very recently, ferromagnetism and superconductivity have been reported to coexist in $CeO_{1-x}F_xBiS_2$ and $Sr_{1-x}Ce_xFBiS_2$ with $T_c$ ~ 2.5-4 K and $T_{FM}$ ~ 4-8 K[8–13]. These materials have layered structure. Magnetism originates in the Ce−O (or Sr/Ce−F) layers and conduction occurs in Bi−$S_2$ layers. In $Sr_{1-x}Ce_xFBiS_2$, which are the parent materials for our Se–added materials $Sr_{1-x}Ce_xFBiS_{2-x}Se_x$, Ce–4*f* electrons provide conduction as well as give rise to long range magnetic order[11]. Ferromagnetic order takes place at a higher temperature and superconductivity sets in an already ferromagnetically ordered lattice at lower temperature. We report here the effect of substitution of *larger isovalent Se* ion at the S site on the magnetic and superconducting properties of $Sr_{1-x}Ce_xFBiS_2$. Se–doping leads to a modest *enhancement* of $T_c$ (upto 3.3 K) and a significant *depression* of $T_{FM}$ (down to 3.5 K). Thus the effect of Se–doping is to move $T_c$ and $T_{FM}$ in *opposite* directions, and thereby bringing them in closer proximity in temperature. We present here results of our investigations of the two materials $Sr_{0.5}Ce_{0.5}FBiS_{1.5}Se_{0.5}$ and $Sr_{0.5}Ce_{0.5}FBiS_{1.0}Se_{1.0}$.

**Experimental**

Polycrystalline compounds of the series $Sr_{0.5}Ce_{0.5}FBiS_{2-x}Se_x$ ($x$ = 0.0, 0.5 and 1.0) were prepared by the usual solid state synthesis procedure as reported elsewhere [14,15]. Phase purity of all the compositions was checked by powder X–ray diffraction technique using Cu−K$\alpha$ radiation source. Temperature dependent resistivity and specific heat measurements were performed using a 14 T PPMS (Quantum Design) and the magnetic properties of the samples were measured using a MPMS7 SQUID magnetometer (Quantum Design).

**Results and Discussion**

Powder X−ray diffraction patterns of all the above mentioned compositions are shown in figure 1. All the peaks could be easily indexed on the basis of a primitive tetragonal unit cell (SG: *P*4/*nmm*). A few minor peaks corresponding to the impurity $Bi_2Se_3$ were also observed. The lattice parameters *a* and *c* obtained by least squares fit method, as expected, show a gradual increase upon Se doping resulting in the unit cell expansion (inset in figure 1). Results of SEM-EDAX studies are shown in figure S1 in supplementary material (SM). Compositional analysis gives a stoichiometry close to the nominal value for both the compositions ($x$ = 0.5 and 1). For the $x$ = 1.0 sample, the Se:S ratio was slightly less than 1. This was possibly due to the formation of traces of an impurity phase $Bi_2Se_3$ (insulator (topological) under ambient pressure)[16] as inferred from our X−ray diffraction patterns of the samples.

Below we describe superconducting and magnetic properties of $Sr_{0.5}Ce_{0.5}FBiS_{2-x}Se_x$ ($x$ = 0.5, 1.0)

**(a) Resistivity**

Resistivity of the materials, measured as a function of temperature is shown in Fig. 2. In the normal state, resistivity of $Sr_{0.5}Ce_{0.5}FBiS_2$ ($x$ = 0) shows semiconducting–like temperature dependence, namely, increase in resistivity with the decrease of temperature just before the onset of superconducting transition at 2.4 K as shown in Fig. 2(a). In the Se–doped materials, $x$ = 0.5 and $x$ = 1.0, this semiconducting behavior is progressively subdued and metallic conductivity is observed in the normal state. Superconductivity sets in at $T_c$ = 2.9 and 3.3 K in materials with $x$ = 0.5 and 1.0 respectively. Our estimate of $T_c^{onset}$ is based on a 90% criterion as shown in figure 2(b). Se–doping clearly *enhances* $T_c$ by ~ 0.5 K in each case (inset of figure 2(b)). In the material with $x$ = 1.0, a sharp superconducting transition is observed with a transition width $\Delta T_c$ = 0.2 K. Similar small enhancement in $T_c$ with Se substitution has been previously observed in $LnO_{1-x}F_xBiS_2$ (Ln = La and Ce) [17–19]. Temperature dependence of upper critical field for $Sr_{0.5}Ce_{0.5}FBiS_{2-x}Se_x$ ($x$ = 0.5 & 1.0), defined at $R = 0.9\ R_N$, is shown in the inset of figure 2(a). For both $x$ = 0.5 and 1.0, the standard WHH model fits well the $B_{c2}(T)$ data. $B_{c2}(0)$ at $T$ = 0 is estimated to be 2.6 T for $x$ = 0.5 and 3.3 T for $x$ = 1.0. These $B_{c2}$ values are at least twice higher than those reported for the Se–free sample $Sr_{0.5}Ln_{0.5}FBiS_2$[11,20]. Enhancement of $T_c$ and $B_{c2}$ in the Se–doped samples is a clear indication that Se atoms have entered the lattice. Se–doping enhances the upper critical field, and therefore reduces the coherence length, of $Sr_{0.5}Ce_{0.5}FBiS_2$

**(b) Magnetic susceptibility in low field of 10 Oe:**

Figure 3 shows dc susceptibility in both the field-cooled (FC) and the zero field-cooled (ZFC) conditions in a field of 10 Oe. Clear diamagnetic signal, of magnitude close to the theoretical value, for both the $x$ = 0.5 and 1.0 compositions is observed in ZFC condition establishing the superconducting state. Poor Meissner response in both cases is possibly due to flux pinning. A superconducting volume fraction of > 90% is estimated for both $x$ = 0.5 and 1.0 compositions suggesting bulk superconductivity in the materials. As deduced from these measurements, superconducting transition temperature increases from $T_c^{onset}$ = 2.65 K for $x$ = 0.5 to $T_c^{onset}$ = 3.10 K for $x$ = 1 which corroborates well with the resistivity data described above. It must be pointed out that in the earlier measurements on Se−free samples $Sr_{0.5}Ce_{0.5}FBiS_2$[11,21] and $CeO_{0.5}F_{0.5}BiS_2$[8,9,13] diamagnetic signal was **not** observed and the occurrence of superconductivity was inferred from the resistivity measurements only.

Further, in Fig. 3, a weak magnetic anomaly is discernible at 3.5 K for the sample $x$ = 0.5 which corresponds to a ferromagnetic transition as evidenced in our high field measurements, as we shall see below, for both the samples $x$ = 0.5 and $x$ = 1.0. This anomaly is not noticeable for the sample $x$ = 1.0 in these low field measurements. $T_c$ and $T_{FM}$ were ascertained from the derivative plots of susceptibility (figure S2 in SM).

**(c) High field DC magnetization measurements:**

Temperature dependence of the magnetic susceptibility $\chi(T)$, measured in an applied field of 10 kOe, and its inverse for the sample $Sr_{0.5}Ce_{0.5}FBiS_{1.5}Se_{0.5}$ is presented in the inset of Figure 3. By fitting the data to the Curie–Weiss law $\chi(T) = \chi_o + C/(T-\theta)$, the paramagnetic effective magnetic moments obtained for the two samples are: $\mu_{eff}$ = 2.22 $\mu_B$ for $x$ = 0.5 and 2.29 $\mu_B$ for $x$ = 1.0 samples (see fig. S3 in SM). These values are close to the theoretical value 2.54 $\mu_B$ for free $Ce^{3+}$ ions. Thus Ce–ions are in trivalent (or nearly trivalent state) state.

We display in Figs. 4a and 4b results of our magnetization measurements, at a few selected temperatures 5 K, 3.5 K and 2 K, as a function of applied magnetic field for the sample $Sr_{0.5}Ce_{0.5}FBiS_{1.5}Se_{0.5}$. At 5 K, magnetization M varies linearly with applied magnetic field, suggesting a paramagnetic state (no magnetic order). At 3.5 K, M is no longer linear in $H$ in the low field region and shows a sign of a ferromagnetic behavior. Similar results have been obtained for $x = 1.0$ composition (shown in fig. S4 in SM). Ferromagnetism state is clearly observed at a lower temperature 2 K and, *remarkably, at this temperature in both the samples, we observe a ferromagnetic hysteresis loop and a* ***superimposed*** *superconducting hysteresis loop*, demonstrating unambiguously the coexistence of ferromagnetism and superconductivity. Observation of such a dual loop is a novel feature of this material, not observed earlier in any superconducting ferromagnet reported to date. In UCoGe, ferromagnetic hysteresis observed in the ferromagnetic state ($T_c < T < T_{FM}$) *gets modified* in the superconducting state ($T < T_c$) but no superconducting hysteresis loop as such was observed [22]. In the virgin low field region, the diamagnetic Meissner response is clearly seen (inset of Fig. 4b) from which $H_{c1}$ is easily estimated to be ~ 50 Oe. Spontaneous vortex state becomes a distinct possibility if the internal field is higher than $H_{c1}$. It is important to point out that in the non–selenium compound $Sr_{0.5}Ce_{0.5}FBiS_2$ ($T_c$ ~ 2.6 K & $T_{FM}$ ~ 7.5 K)[11,21] no superconducting hysteresis loop was observed at 2 K which may be regarded as the parent of the materials under investigation. In a similar material $Ce(O,F)BiS_2$ ($T_c$ ~ 2.5–4 K & $T_{FM}$ ~ 6.5–7.5 K) [9,12,13,23] also no superconducting hysteresis loop was observed. This gives a strong evidence of the crucial changes created by Se–doping in the superconducting and magnetic properties of $Sr_{0.5}Ce_{0.5}FBiS_2$, the parent material.

Dual hysteresis loop has been observed very recently in $[(Li_{1-x}Fe_x)OH](Fe_{1-y}Li_y)Se$[24]. However, this case is different from our case in one crucial sense, namely, in this case, $T_c$ (~ 43 K) >> $T_{FM}$ (10 K) whereas in our case, $T_{FM} > T_c$ and, hence, superconductivity sets in an already ferromagnetically ordered lattice. Further, in our case, superconductivity appears just *at the border of ferromagnetic transition* ($T_{FM}$ is only marginally higher than $T_c$) whereas in the above–mentioned material, superconductivity and ferromagnetism are far separated in temperature. Also, in our case Ce–*4f* electrons are responsible both for magnetism and superconductivity whereas in the $[(Li_{1-x}Fe_x)OH](Fe_{1-y}Li_y)Se$, Fe–3*d* electrons carry both, magnetism and superconductivity. Coexistence of superconductivity and ferromagnetism (with $T_c > T_{FM}$) in $CeFeAs_{1-x}P_xO_{0.95}F_{0.05}$ in limited doping range has been observed. In this case however Ce carries full moment and the system is not itinerant ferromagnet.

The spontaneous magnetization $M_s$ is estimated by linear extrapolation of the high–field data to $H = 0$ (fig. 4a). From the estimated $M_s$, we obtain at $T = 2$ K, the ordered spontaneous Ce–moment $\mu_0$ ~0.09 $\mu_B$ for the sample $x = 0.5$ and 0.11 $\mu_B$ for the sample $x = 1.0$. These values are quite small as compared with what is expected for free $Ce^{3+}$ ion. We may note here that in $Ce(O,F)BiS_2$ a reduced moment $M_s = 0.52$ $\mu_B$/Ce was reported[9] which, possibly, suggests that in this case Ce–ions may be in the crystal–field split doublet state (localized moment). In our case, we observe a drastically reduced, but non–zero, Ce–moment.

It is particularly remarkable that the transition of Ce from the high paramagnetic effective moment $\mu_{eff}$ ~ 2.5$\mu_B$ to a small moment $\mu_0$ ~ 0.1$\mu_B$ in the superconducting state takes place in a tiny interval of temperature, from ~ 3.5 K to ~ 3.1 K. The *high* ratio $\mu_{eff}/\mu_0$ (~ 25) implies an itinerant ferromagnetic state in both materials $Sr_{0.5}Ce_{0.5}FBiS_{2-x}Se_x$, $x$ = 0.5 and $x$ = 1.0 [7,25]. Among the U–containing superconducting ferromagnets[26] UCoGe[27] has the highest superconducting transition temperature (~ 0.6 K), highest ferromagnetic ordering temperature ($T_{FM}$ ~ 3 K) at ambient pressure and the itinerant magnetic moment in the coexisting superconducting state of about 0.03–0.07 $\mu_B$. UCoGe has been argued to be a *p*–wave superconductor, thus, we are inclined to propose that our materials are also *p*–wave superconductors. Clearly more work is required to establish the *p*-wave pairing in these materials. These materials fill the glaring void, namely, so far no Ce-based material has been hitherto known exhibiting superconductivity within the itinerant ferromagnetic state. The present Ce–based materials have better superconducting and magnetic characteristics.

**(d) Specific heat**

Figure 5 shows the temperature dependence of specific heat of $Sr_{0.5}Ce_{0.5}FBiS_{1.5}Se_{0.5}$ ($x$ = 0.5) in the low temperature range 2−16 K. Inset shows C/$T$ data before (blue circle) and after subtraction (black circle) of a Schottky contribution which was approximated by the dashed line. A broad peak, not $\lambda$–shaped as expected for a ferromagnet, centered at 3.2 K (inset of Fig. 5) is observed from which, following Li *et al* [11], ferromagnetic ordering temperature $T_{FM}$ ~ 3.6 K is obtained. As $T_{FM}$ and $T_c$ are quite close, the anomalies of the two transitions, the magnetic and the superconducting, are not resolved. The red line in the main panel obeys the equation: $C/T = \gamma + \beta T^2$ yielding the Debye temperature $\theta_D$ = 187 K and the Sommerfeld coefficient $\gamma$ = 12 mJ/(K$^2$mol Ce). This $\gamma$ value of $Sr_{0.5}Ce_{0.5}FBiS_{1.5}Se_{0.5}$ is much smaller than that of $Sr_{0.5}Ce_{0.5}FBiS_2$ ($\gamma$ = 117 mJ/(K$^2$mol Ce)) [11]. However, it is increased by a factor of 5−10 as compared to $Sr_{1-}$ $_x$La$_x$FBiS$_2$ ($\gamma$< 2 mJ/mol-K$^2$) [20,28,29] and La$_{1-x}$M$_x$OBiS$_2$ (M =Ti, Zr, Th) ($\gamma$ ~ 0.58–2.21 mJ/mol K$^2$) [30]. Therefore, we suppose that 4*f* electrons of Ce contribute to the density of states at the Fermi level and $Sr_{0.5}Ce_{0.5}FBiS_{1.5}Se_{0.5}$ is rather strongly correlated system compared to other $BiS_2$–based materials. Se doping may reduce hybridization between itinerant Bi–6*p* electrons and heavy 4*f* electrons leading to stronger localization of Ce moments. Therefore, we attribute the higher $\gamma$ value for $Sr_{0.5}Ce_{0.5}FBiS_2$ to the electronic correlation effect of Ce–4*f* electrons which is much reduced in the Se–doped sample. From the specific heat measurements we get an entropy per Ce atom of only about 4 % of the expected value for J = 5/2. The low magnetic entropy ($S_m$ = 0.04*R*ln6) is consistent with weak itinerant ferromagnetism in $Sr_{0.5}Ce_{0.5}FBiS_{1.5}Se_{0.5}$. This situation is similar to that in UCoGe[7].

Though no superconducting ferromagnets based on Ce are known hitherto, there are, however, well known superconducting antiferromagnets such as $CeRhIn_5$ [31,32]. Very recently $CePt_2In_7$ has been shown to be an interesting superconducting antiferromagnet [33]. Thus $Sr_{0.5}Ce_{0.5}FBiS_{2-x}Se_x$ is an important and timely addition to the exciting Ce–based materials exhibiting coexisting superconductivity and magnetism

**Concluding Remarks**

We have observed superconductivity ($T_c$ ~ 3.0 K) and itinerant ferromagnetism ($T_{FM}$ ~ 3.5 K) *coexisting* in the new materials $Sr_{0.5}Ce_{0.5}FBiS_{2-x}Se_x$. Thus in these materials, superconductivity occurs much closer to the border of ferromagnetism than in UCoGe, the most interesting U–based superconducting ferromagnet. A novel feature of these materials *is a **dual** hysteresis loop corresponding to **both** the coexisting superconductivity and ferromagnetism.* These materials, with such favourable characteristics, appear to have an edge over UCoGe. They are potential candidates for the unconventional *p*–wave superconductivity which deserves to be further persued. We are making efforts to grow single crystals of these materials and if we succeed, we would carry out studies such as NMR, MuSR and neutron diffraction and Andreev reflection to throw light on the nature of the superconducting state in these materials.

**Corresponding author**:

Ashok K. Ganguli

Department of Chemistry

Indian Institute of Technology, New Delhi, India, 110016

Phone: 011 2659 1511

Email: ashok@chemistry.iitd.ac.in

**Acknowledgements:**

Author AKG thanks DST (Govt. of India) for partial financial assistance. GST and ZH thank CSIR and UGC (Govt. of India) respectively for fellowships. Authors at IIT Delhi thank PPMS EVERCOOL-II facility at IIT Delhi.

**Notes and References**

[‡] Visiting scientist at Solid State and Nano Research laboratory, Dept. of Chemistry, IIT Delhi.

**Figure 1.** Powder X-ray diffraction of $Sr_{0.5}Ce_{0.5}FBiS_{2-x}Se_x$ ($x$ = 0.0. 0.5 and 1.0). Asterisk marks the impurity phase $Bi_2Se_3$. Inset shows the variation of cell volume with Se content.

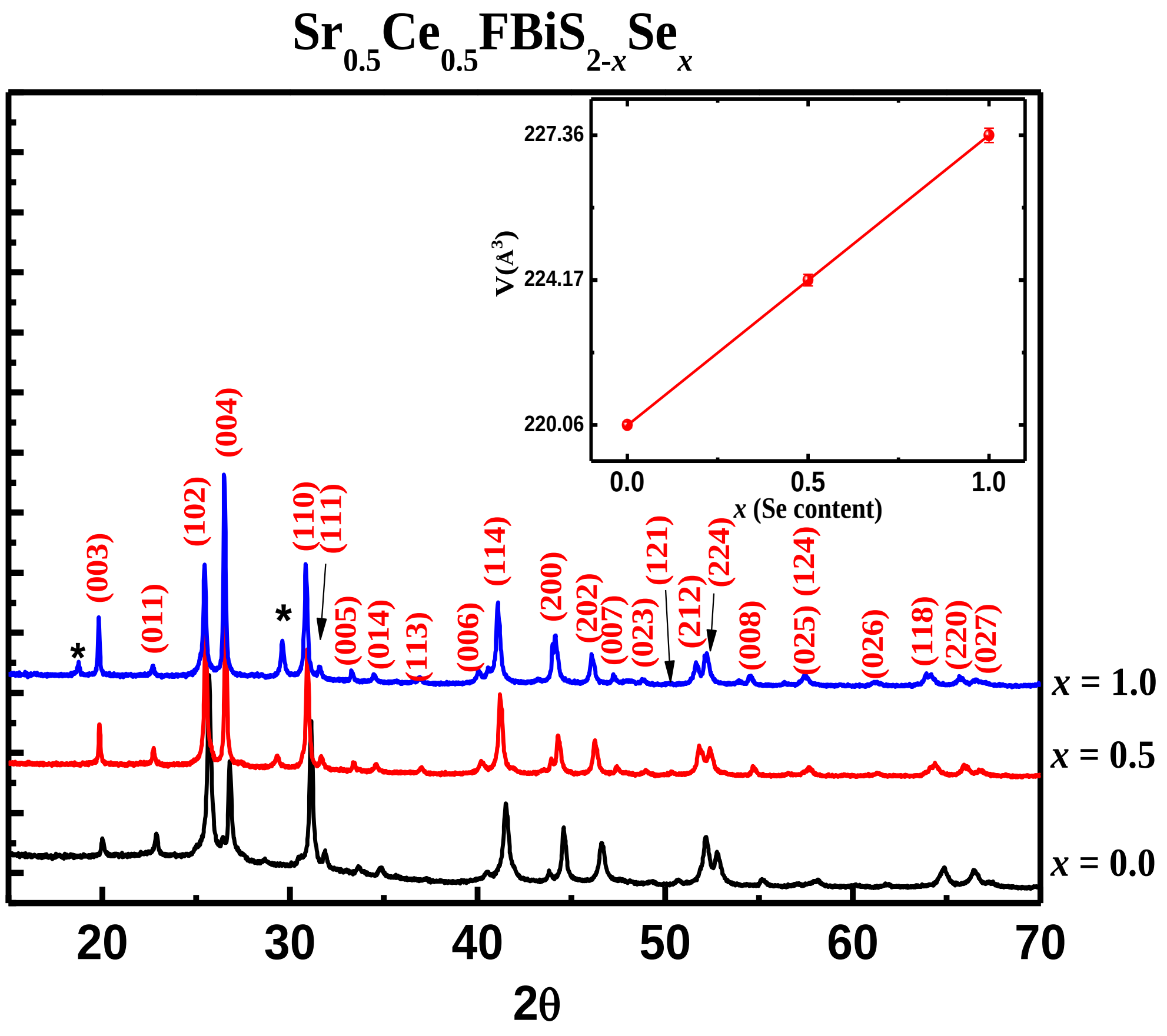

**Figure 2.** Variable temperature resistivity curves for $Sr_{0.5}Ce_{0.5}FBiS_{2-x}Se_x$; $x$ = 0, 0.5 and 1.0 (a) in temperature range 2–300 K and (b) in low temperature range. Notice both the materials show a clean resistive transition with a small width (ΔT = 0.2 K). Inset of (a) shows the upper critical field ($B_{c2}$) versus temperature (T) curve for the $x$ = 0.5 and 1.0 compositions (open circles) along with the WHH fit (solid lines). Inset of (b) show the variation of $T_c^{onset}$ and $T_c$ (ρ = zero) as a function of Se-doping.

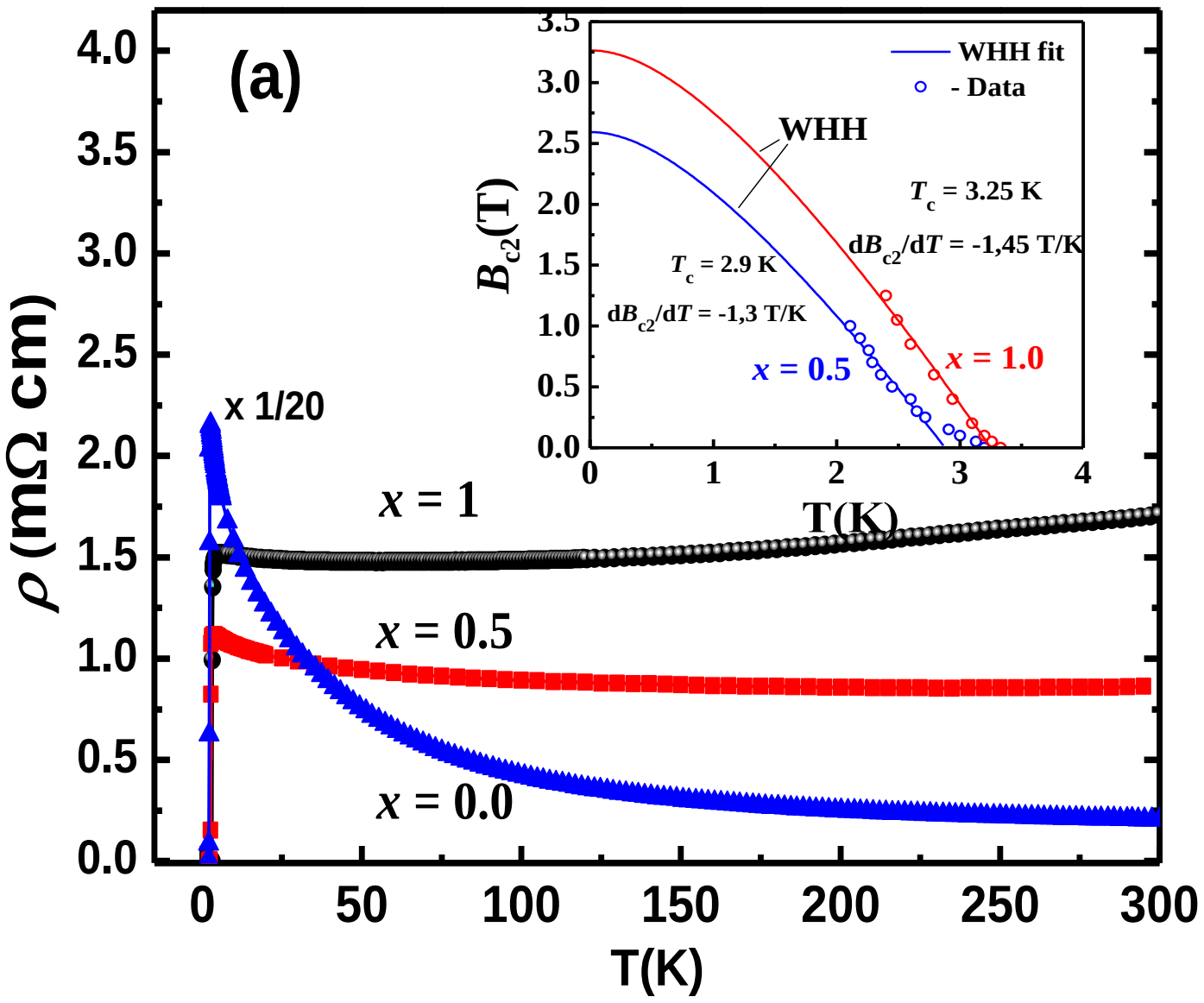

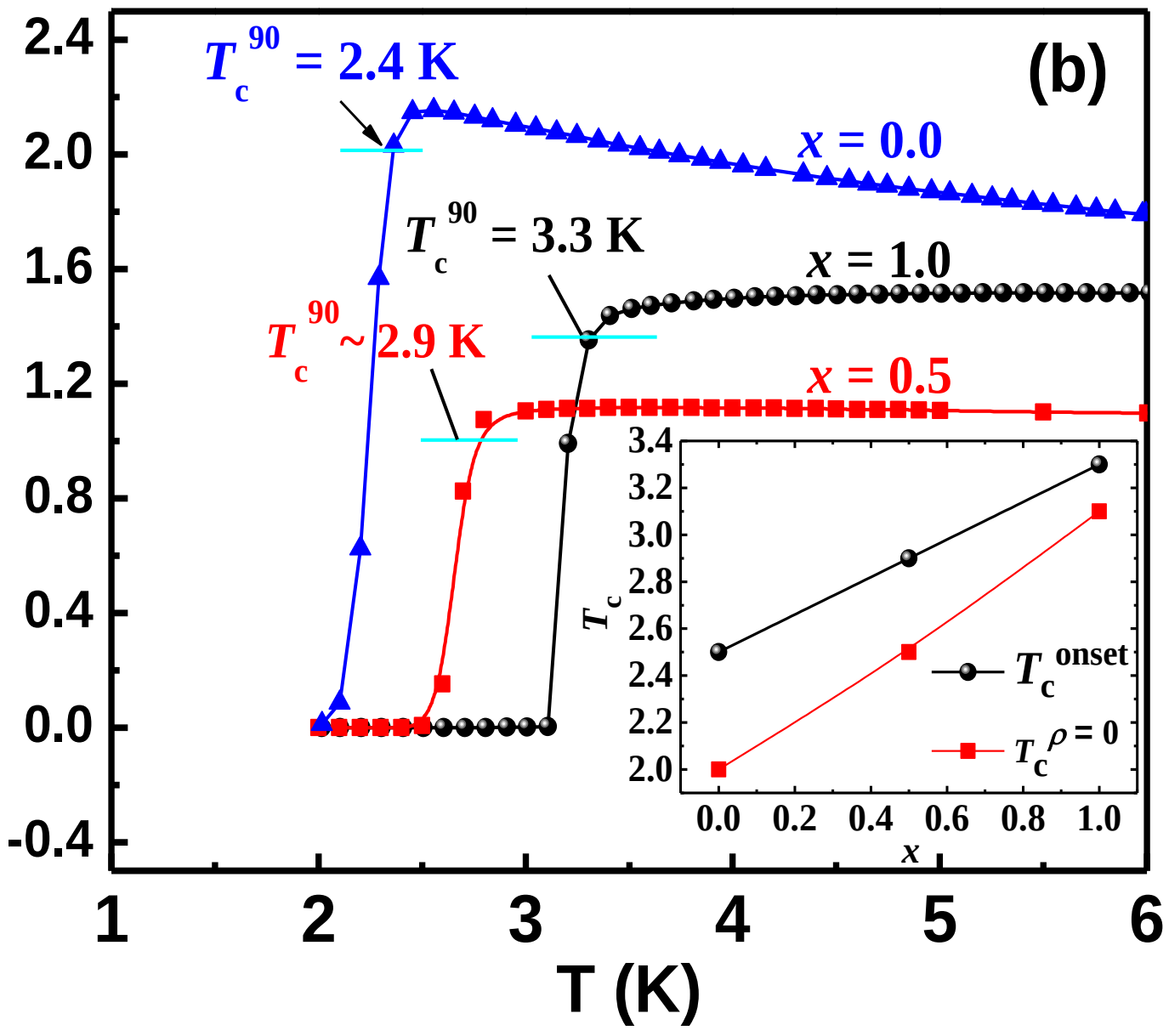

**Figure 3.** Variable temperature *dc* susceptibility of the $Sr_{0.5}Ce_{0.5}FBiS_{2-x}Se_x$ ($x$ = 0.5, 1.0) in an applied field of 10 Oe. Inset show paramagnetic susceptibility (black) and its inverse (green). Red line is the Curie–Weiss fit.

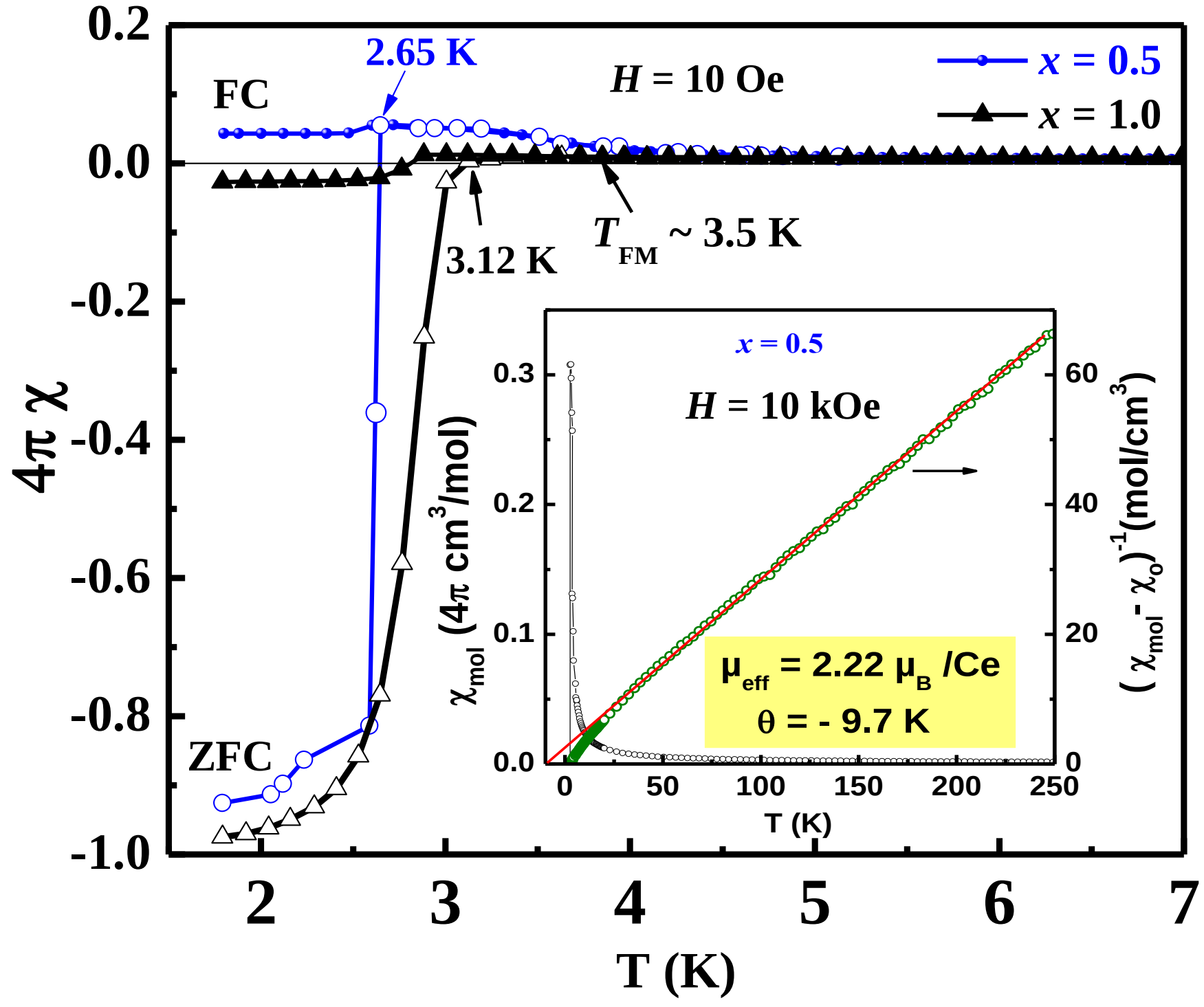

**Figure 4.** Hysteresis loops at different temperatures for $Sr_{0.5}Ce_{0.5}FBiS_{1.5}Se_{0.5}$ in high field range (**a**) and low field range (**b**). The superconducting loop is superimposed on ferromagnetic loop at 2 K. Inset of (b) shows the initial diamagnetic signal.

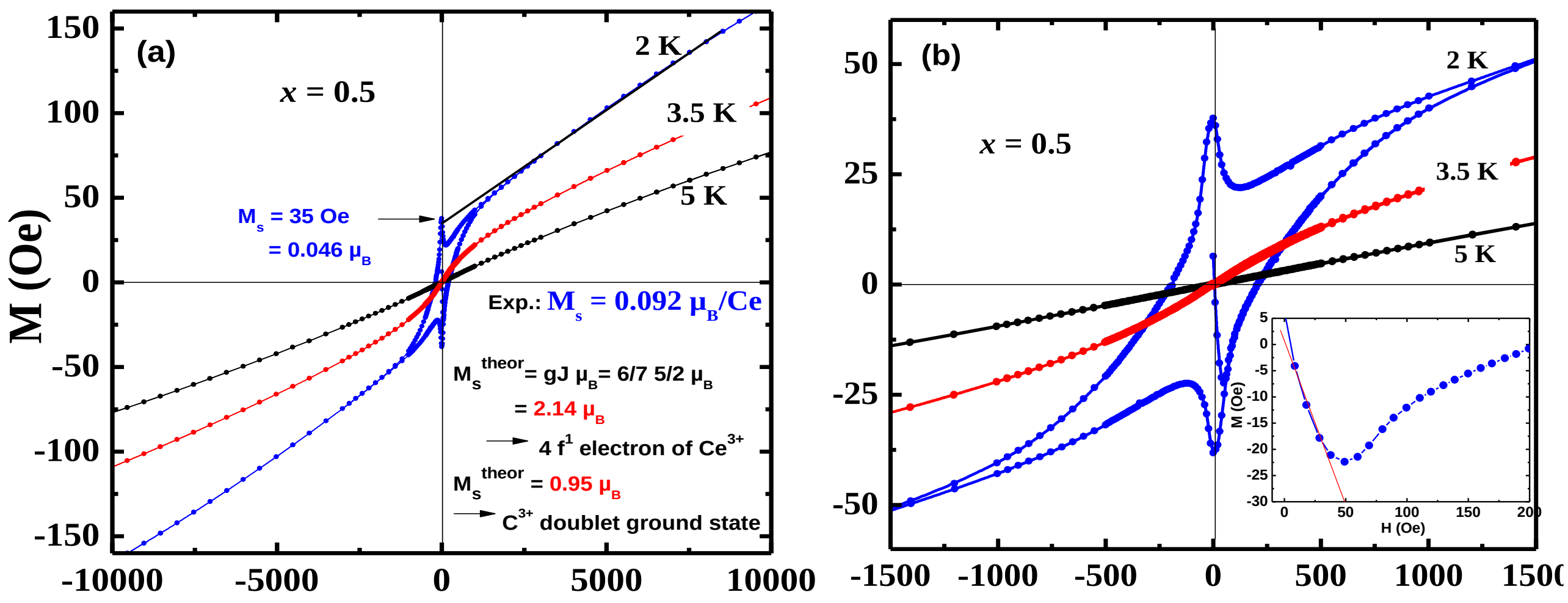

**Figure 5.** Temperature dependence of Schottky corrected specific heat C/$T$ vs.$T^2$ for $x$ = 0.5 sample at $H$ = 0 and low $T$. Red line is the linear fit in the equation C/T = $\gamma + \beta T^2$. Inset show C/$T$ data before (blue circle) and after subtraction (black circle) of a Schottky contribution which was approximated by the dashed line.

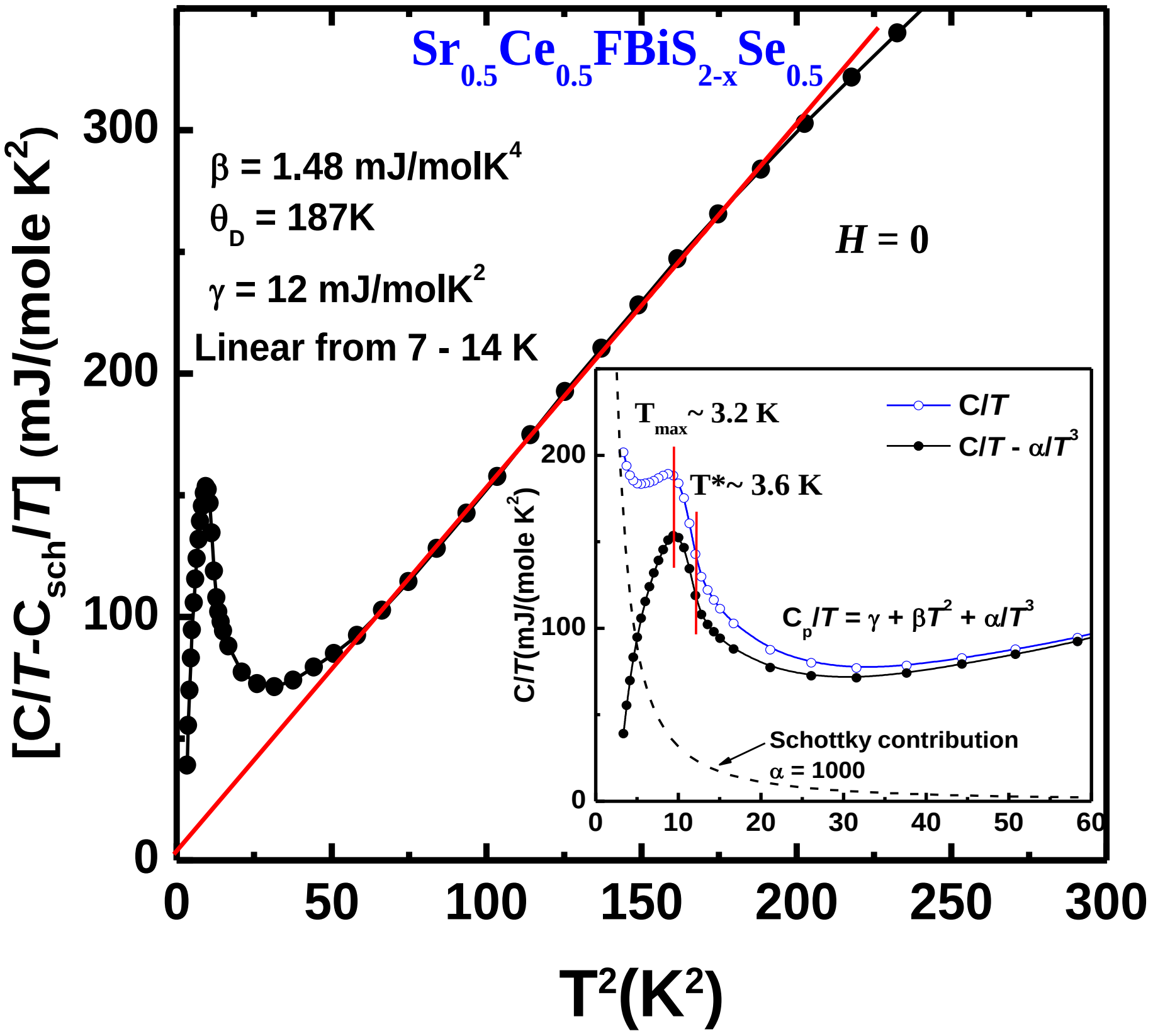

**Supplementary Materials**

**Figure S1.** Compositional analysis of $Sr_{0.5}Ce_{0.5}FBiS_{2-x}Se_x$ ($x$ = 0.5, 1.0) samples by SEM–EDAX. The typical spectra show the presence of all the constituent elements in the selected region (inset). Many such regions were selected and all the monitored regions had acceptable S and Se contents, thus eliminating the possibility of any phase separation.

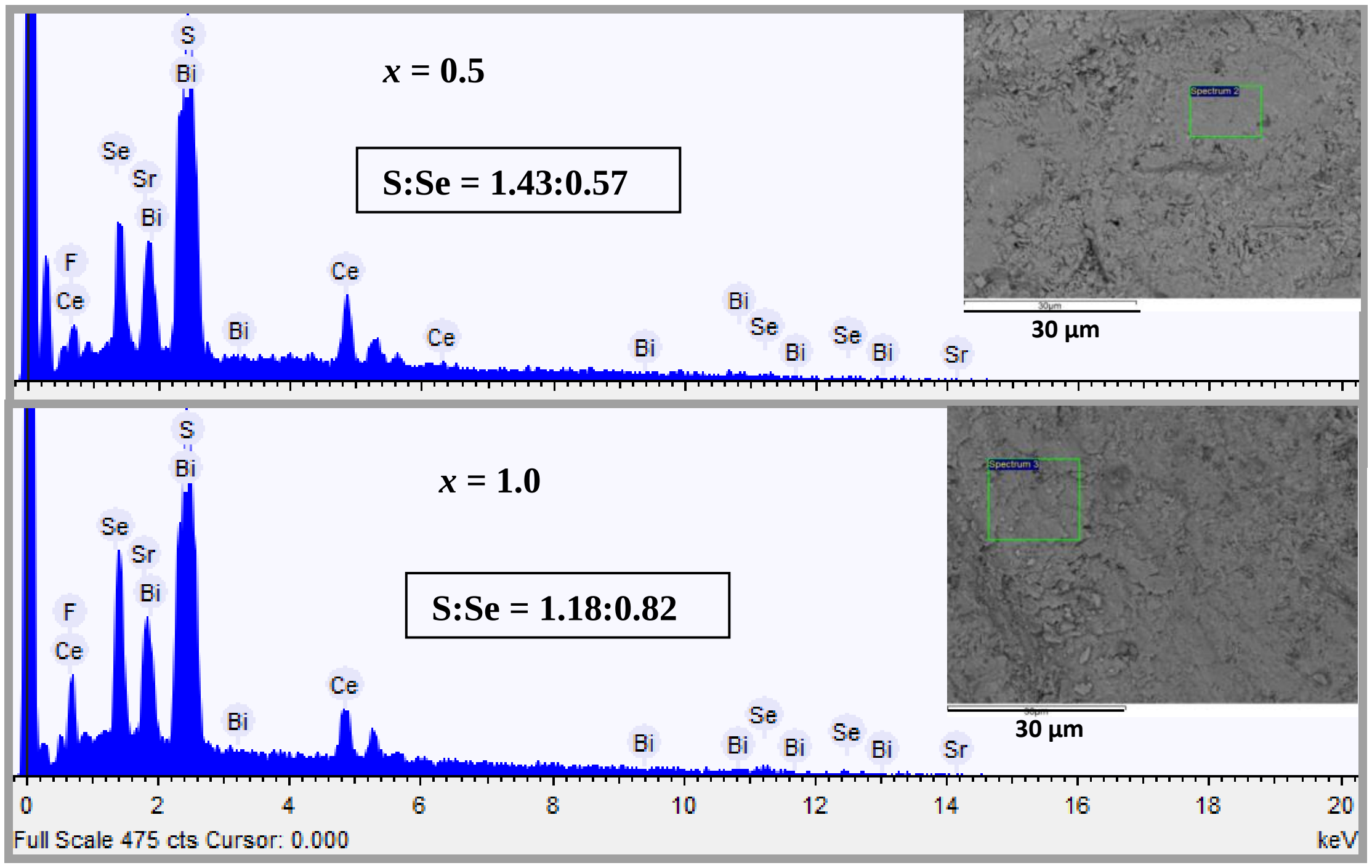

**Figure S2.** Temperature dependent magnetic susceptibility and its derivative indicating the $T_c$ and $T_{FM}$ for $Sr_{0.5}Ce_{0.5}FBiS_{2-x}Se_x$ (a) $x$ = 0.5 and (b) $x$ = 0.1.

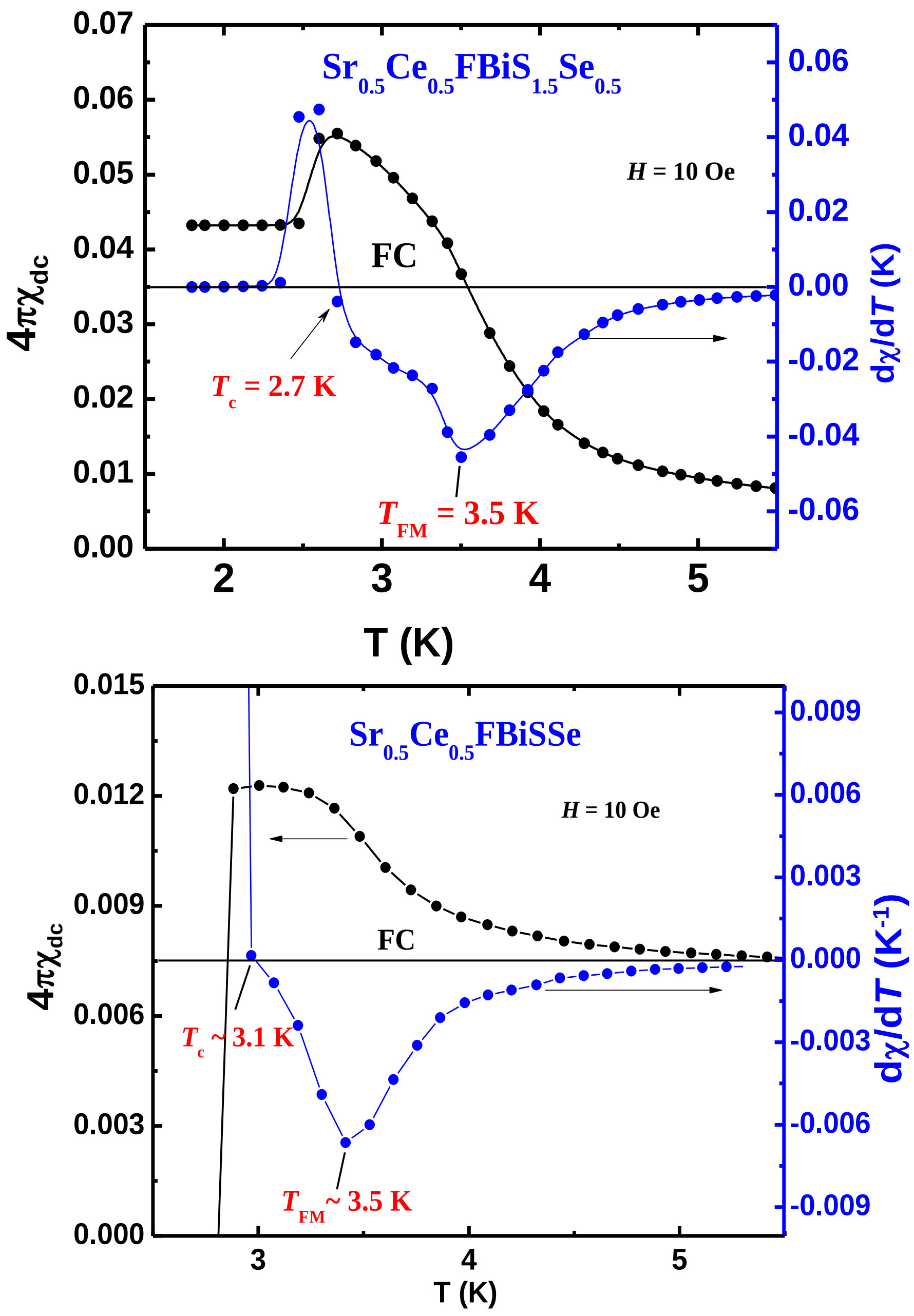

**Figure S3.** Variable temperature paramagnetic susceptibility (black) and its inverse (green) for $Sr_{0.5}Ce_{0.5}FBiS_{1.0}Se_{1.0}$. Red line is the Curie–Weiss fit.

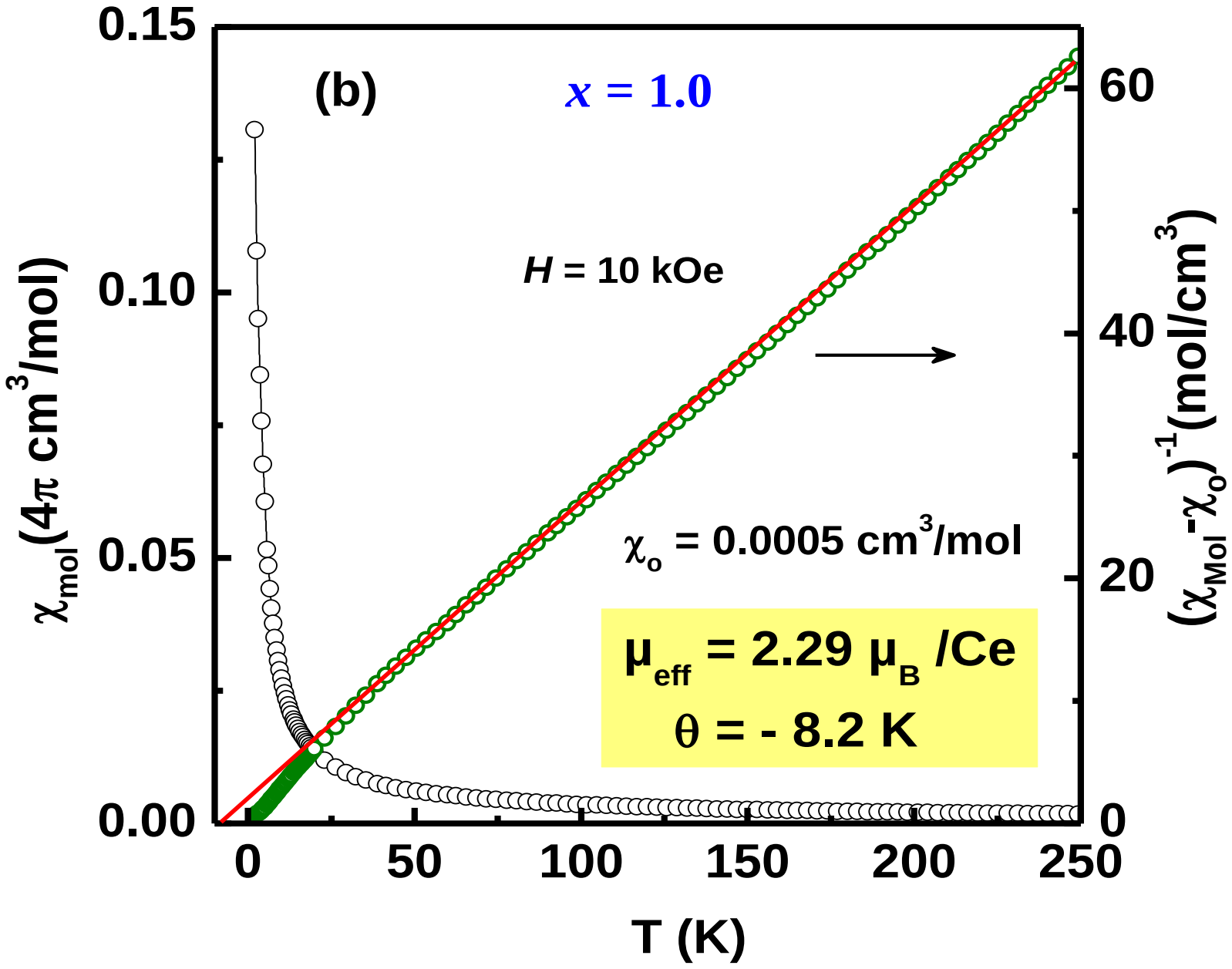

**Figure S4.** Hysteresis loops at different temperatures for $x$ = 1.0 sample in high field range (**a**) and low field range (**b**). The superconducting loops are superimposed on ferromagnetic loops at 2 K for both the samples. Inset of (b) show the initial diamagnetic signal.

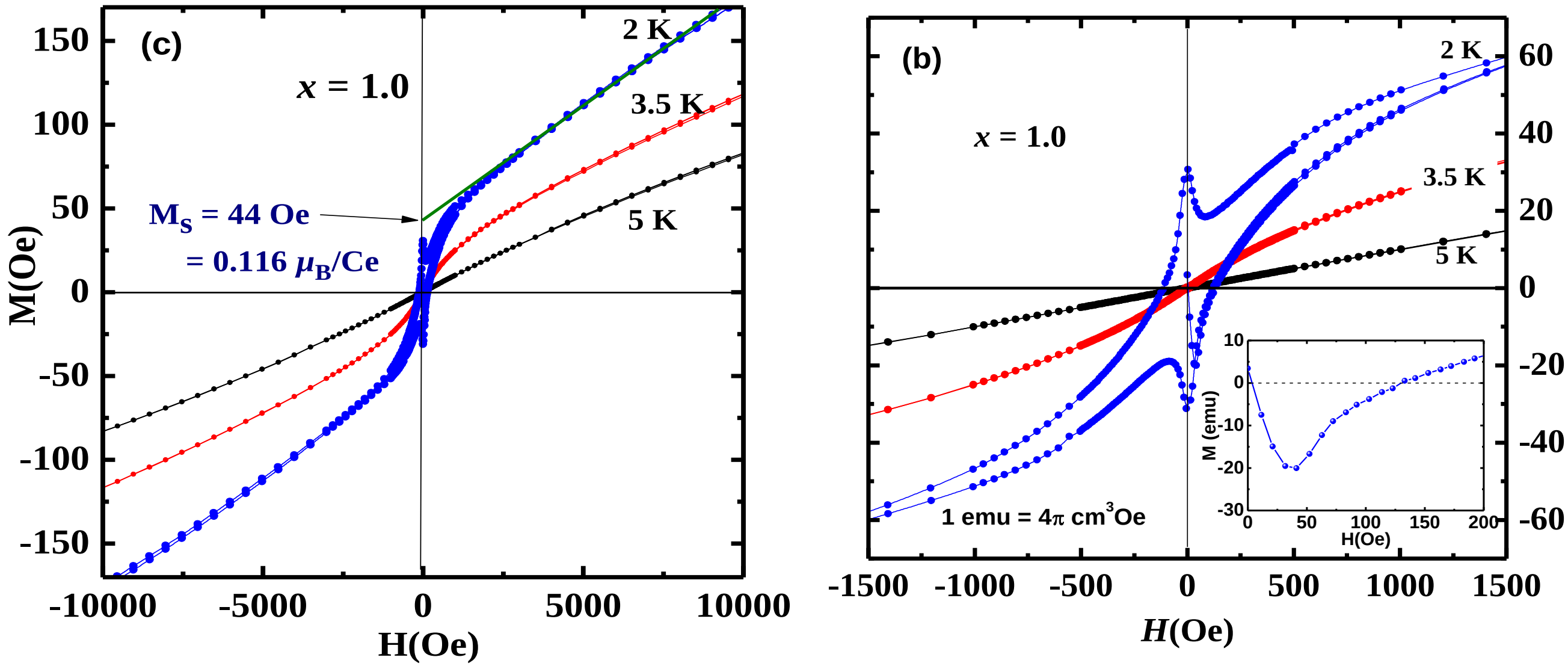